\documentclass{acm_proc_article-sp}
\usepackage{graphicx}
\usepackage{url}

\usepackage{algpseudocode}
\usepackage{algorithm}
\begin{document}

\title{LabelRankT: Incremental Community Detection in Dynamic Networks via Label Propagation}
%\subtitle{[Extended Abstract]
%\titlenote{A full version of this paper is available as
%\textit{Author's Guide to Preparing ACM SIG Proceedings Using
%\LaTeX$2_\epsilon$\ and BibTeX} at

\numberofauthors{3} %  in this sample file, there are a *total*
\author{
% 1st. author
\alignauthor
Jierui Xie\\
       \affaddr{Rensselaer Polytechnic Institute}\\
       \affaddr{Troy, New York 12180, USA}\\
       \email{jierui.xie@gmail.com}
% 2nd. author
\alignauthor
Mingming Chen\\
       \affaddr{Rensselaer Polytechnic Institute}\\
       \affaddr{Troy, New York 12180, USA}\\
       \email{chenm8@rpi.edu}
\alignauthor
Boleslaw K. Szymanski\\
       \affaddr{Rensselaer Polytechnic Institute}\\
       \affaddr{Troy, New York 12180, USA}\\
       \email{szymab@rpi.edu}
}
\maketitle
\begin{abstract}
An increasingly important challenge in network analysis is efficient detection and tracking of communities in dynamic networks for which changes arrive as a stream. There is a need for algorithms that can \textit{incrementally} update and monitor communities whose evolution generates huge real-time data streams, such as the Internet or on-line social networks. In this paper, we propose \textit{LabelRankT}, an on-line distributed algorithm for detection of communities in large-scale dynamic networks through stabilized label propagation. Results of tests on real-world networks demonstrate that \textit{LabelRankT} has much lower computational costs than other algorithms. It also improves the quality of the detected communities compared to dynamic detection methods and matches the quality achieved by static detection approaches. Unlike most of other algorithms which apply only to binary networks, \textit{LabelRankT} works on weighted and directed networks, which provides a flexible and promising solution for real-world applications.
\end{abstract}

%A category including the fourth, optional field follows...
%\category{D.2.8}{Software Engineering}{Metrics}[complexity measures, performance measures]


\keywords{social network, community detection, clustering, network evolution, dynamic network, temporal} % NOT required for Proceedings

\section{Introduction}
{\it Communities} are the basic structures in sociology in general and in social networks in particular. They have been intensively researched for more than a half of the century~\cite{human-comm}. 
In sociology, community usually refers to a social unit whose members share common 
values and the identity of the members as well as their degree of cohesiveness depend on individuals' social and cognitive factors such as beliefs, 
preferences, or needs. The ubiquity of the Internet and social media eliminated spatial limitations on community range, enabling on-line communities to link people regardless of their physical location. 
The newly arising {\it computational sociology} relies on computationally intensive methods to analyze and model social phenomena \cite{comput-soc}, including communities and their detection. 

Analysis of social networks became one of the basic tools of sociology \cite{WSFK94} and
%Social network analysis: Methods and applications}, Wasserman, Stanley and Faust, Katherine}, volume={8}, year={1994}, publisher={Cambridge university press}
has been used for linking micro and macro levels of 
sociological theory. The classical example of the approach is presented in \cite{weak-ties} that elaborated the macro implications of one aspect of small-scale interaction, the strength of dyadic ties.
%However, as shown in \cite{Cosma Rohilla Shalizi and Andrew C. Thomas, Homophily and Contagion Are Generically Confounded in  Observational Social Network Studies, Sociological Methods & Research 2011 40: 211}, uniquely matching such interactions to social factors is usually not possible. For example, homophily, social congestion, and individual traits are confounded with each other. Distinguishing between them requires strong assumptions on the parameterization of the social process or on the adequacy of the covariates used (or both).   
Moreover, a lot of commercial applications, such as digital marketing, behavioral targeting and user preference mining, rely heavily on community analysis. 

With the rapid growth of large-scale on-line social networks, e.g., Facebook connected a billion users in 2012, there is a high demand for efficient community detection algorithms that will be able to handle their evolution growth. Communities in on-line social networks are discovered by analyzing the observed and often recorded on-line interactions between people. 

Numerous techniques have been developed for community detection. However, most of them require a \textit{global} and often \textit{static} view of the network and ignore temporal correlations between different snapshots over time. Such algorithms are not scalable enough to cope with dynamically evolving networks, especially when new data about them are generated continuously. Another limitation shared by most of the existing algorithms is that they are applicable only to networks with binary adjacency matrix, that is with undirected and unweighted edges.

Label propagation based community detection algorithms such as LPA \cite{Raghavan:2007,JieruiXieLPA:2010}, COPRA \cite{Gregory:2010} and SLPA\footnote{Codes: \url{https://sites.google.com/site/communitydetectionslpa/}} \cite{JieruiXie-SLPA:2012} have been shown to perform well in static networks. However,  due to random tie breaking strategy, they produce different partitions in different runs. Such instability is highly undesirable when tracking the evolution of communities in a dynamic network.

The contributions of this paper are two-fold. First, we generalized the \textit{LabelRank} algorithm introduced in \cite{JieruiXie-LabelRank-NSW:2013} to incorporate important network features such as edge weights and directions. Second, built upon \textit{LabelRank}, we introduce \textit{LabelRankT} algorithm that incrementally detects evolving communities in dynamic networks. The new algorithm presented here delivers significant improvements over the existing solutions
in both the quality of detected evolving communities and the speed of program execution.

\newpage

\section{Generalization of LabelRank}
To make the paper self-contained, this section first summarizes and generalizes the \textit{LabelRank} algorithm introduced in \cite{JieruiXie-LabelRank-NSW:2013} which is the basis for \textit{LabelRanlT}. Both algorithms are based on the idea of simulating the propagation of labels in the network.

\textit{LabelRank} relies on four operators applied to the labels: (i) propagation, (ii) inflation, (iii) cutoff, and (iv) conditional update. The data structure that lies in the core of this algorithm is the sparse matrix of label distribution. Each node maintains a label distribution locally during the propagation. At the end of the algorithm, \textit{LabelRank} ranks labels in each node. Nodes with the same highest probability label form a community.

\textit{LabelRank}, a stabilized LPA, was initially introduced in \cite{JieruiXie-LabelRank-NSW:2013} for binary networks. Here, we generalize the propagation operation over edges in order to take both edge direction and weight into account. Given a network $G=(E,V)$, where $E$ is the set of edges and $V$ is the set of nodes, this operator can be expressed in matrix form as:
\begin{equation}
\label{eq:AP}
W \times P,
\end{equation}

%-------------------------------------------
%\begin{algorithm}[H]
\begin{algorithm}[t]
\caption{Generalized \textit{LabelRank}}
\label{alg1}
\begin{algorithmic}[1]
\State add self-loop to adjacency matrix $W$
\State initialize label distribution $P$ using Eq. \ref{eq:initP}
\Repeat
	\State $P'=W \times P$
	\State $P'= \Gamma _{in} P'$
	\State $P'= \Phi_{r} P'$
	\State $P= \Theta _{q}(P',P)$
\Until{stop criterion satisfied}
\State output communities
\end{algorithmic}
\end{algorithm}

where $W$ is the $n \times n$ weight matrix and $n$ is the number of nodes. For each $w_{ij}$, if there exist a directed edge $e_{ji} \in E$ from node $j$ to $i$, $w_{ij}$ takes on a positive value; otherwise it is 0. $w_{ij}>0$ is the weight placed on the directed edge, which bears important and application specific information. $P$ is the $n \times n$ \textit{label distribution matrix}. Label is just a unique identifier. For simplicity, it usually takes on the same value as the node ID. $P$ is composed of $n$ $1 \times n$ row vectors $P_i$, one for each node $i$. Each element $P_{ic}$ or $P_i(c)$ holds the current estimation of probability of node $i$ observing label $c \in C$ taken from a finite set of alphabet $C$ (here $|C|=n$).

Each node broadcasts the distribution to its neighbors at each iteration step and computes the new distribution $P_i^{'}$ simultaneously using the following equation:
\begin{equation}
\label{eq:bc}
P_{ic}=\frac{\sum_{j\in Nb(i)} w_{ij} P_{jc}}{\sum_{k\in Nb(i)}{w_{ik}}} , \forall c\in C,
\end{equation}
where $Nb(i)$ is a set of neighbors of node $i$ and the numerator sums up weights of all edges incoming node $i$. That is, a node sends out its information along outgoing edges to its neighbors and at the same time receives information along incoming edges from the neighborhood. Note that, $P_i^{'}$ is normalized to make proper probability distribution. One can show that the new distribution vector $P_i^{'}$ is the distribution that minimizes the KL divergence between any possible $P_{i}^{'}$ and $P_{i}$ \cite{Tsuda:2005}.

To initialize $P$, each node is assigned a distribution of probabilities of all incoming edges by assigning to each incoming edge the initial probability of seeing this neighbor's label proportional to the weight of this edge:
\begin{equation}
\label{eq:initP}
P_{ij}=\frac{w_{ij}}{\sum_{k \in Nb(i)}{w_{ik}}}, \forall j\in C \mbox{ s.t. } w_{ij} > 0.
\end{equation}

We briefly characterize the three remaining operations of \textit{LabelRank} as follows.

(1) The inflation operator $\Gamma _{in}$ on $P$ \cite{MCL:2000} is used to contract the propagation, where $in$ is the parameter taking on real values. It operates on the label distribution matrix $P$ (rather than to a stochastic matrix or adjacency matrix) to decouple it from the network structure. After applying $\Gamma _{in} P$, each $P_{ic}$ is proportional to $P_{ic}^{in}$, i.e., $P_{ic}$ rises to the $in^{th}$ power.

(2) The full label propagation distribution induces a cost for memory. To alleviate this problem, the cutoff operator $\Phi_{r}$ on $P$ is introduced to remove labels that are below threshold $r\in [0,1]$. More importantly, $\Phi_{r}$ is shown empirically to reduce the space complexity efficiently, from quadratic to linear. The average number of labels in each node is typically less than $3.0$ for $r=0.1$.

(3) The conditional update $\Theta _{q}$ operator is used to trap the process in the quality space (e.g., modularity \cite{newman-2004-69}) to avoid trivial network state where each node holds the same distribution. The algorithm updates a node only when it is significantly different from its neighbors in terms of labels. This allows us to preserve detected communities and detect termination based on scarcity of changes to the network. At each iteration, the change is accepted only by nodes that satisfy the following update condition:
\begin{equation}
\label{eq:conditon_update}
\sum_{j \in Nb(i)} \textit{isSubset}(C_i^*,C_j^*) \le qk_i,
\end{equation}
where $C_i^*$ is the set of maximum labels which includes labels with the maximum probability at node $i$ at the previous time step. Function $\textit{isSubset}(s_1,s_2)$ returns $1$ if $s_1 \subseteq s_2$, and $0$ otherwise. $k_i$ is the degree of node $i$, and $q$ is a real number parameter chosen from the interval $[0, 1]$. Intuitively, \textit{isSubset} can be viewed as a measure of similarity between two nodes.

These four operators together with a post-processing that groups nodes whose highest probability labels are the same into a community form our algorithm (see Alg.~\ref{alg1}).

%These four operators together with a post-processing that groups nodes whose highest probability labels are the same into a community form a complete algorithm (see Alg.~\ref{alg1}). An example network as output by LabelRank is shown in Fig.~\ref{fig:G0} (with cutoff $r=0.1$, inflation $in=4$, and conditional update $q=0.7$). There are only 12 labels on average and at most two in each node, resulting in a sparse label distribution (Table \ref{table:tabelG0} of which second row shows for each node the label with the highest probability identifying this node community). Three communities are identified, each sharing a common label: red community label 3, green community label 5 and blue community label 11. The resultant $P$ also distinguishes two types of nodes, the \textit{border} ones with high probability labels (e.g., 3, 5 and 11), and the \textit{core} nodes with positive but not largest label probabilities (e.g., 1, 13 and 15). The latter are well connected to their communities.

%-------------------------------------------
%\begin{algorithm}[H]
\begin{algorithm}[t]
\caption{\textit{LabelRankT}}
\label{alg2}
\begin{algorithmic}[1]
\State input: snapshots $G([0, 1, \cdots, T])$
%\State run LabelRank on $G(0)$ to produce $P^0$
\For{t=1:T}
\State 	(a) Tracking the \textit{changed} nodes in $G(t)$ due to the changes in edges they attach to since $t-1$.
\State 	(b) Initialize $P^t$. For node $i$ that does not change since $t-1$, we copy its label distributions, i.e., $P_i^{t}=P_i^{t-1}$ . For \textit{changed} nodes, we reinitialize their label distributions as in \textit{LabelRank}.
\State 	(c) Iteratively update only \textit{changed} nodes' label distribution and assign them to the corresponding communities as in \textit{LabelRank}.	
\EndFor
\end{algorithmic}
\end{algorithm}	

%-------------------------------------------
\begin{figure}[hbtp]	
  \centering
	\includegraphics[scale=0.4]{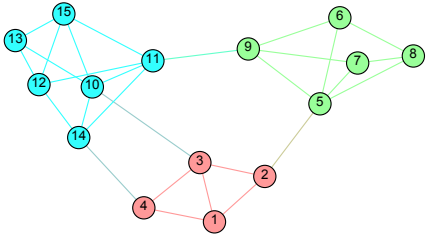}
	\caption{The example network G(0) with $n=15$. Colors represent communities discovered by \textit{LabelRankT}.}
	\label{fig:G0}
\end{figure}

%-------------------------------------------
%side by side 1-2
%\begin{figure*}[htbp]
%-----------------------------------------------

%-----------------------------------------------
\begin{figure}[htbp]
\centering	
	\includegraphics[scale=0.4]{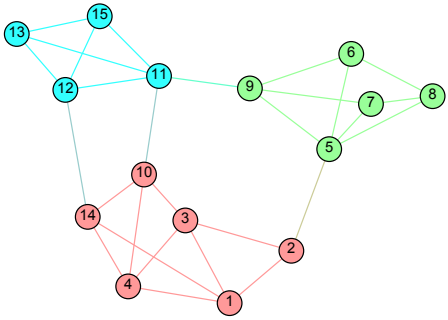}
				\caption{The example network G(1) with n = 15 that change from G(0) by splitting and merging. Nodes 10 and 14 moved from blue group into the red group. Three edges were deleted from and three were added. Colors represent communities discovered by \textit{LabelRankT} after these events.}
				\label{fig:G1}
\end{figure}

\begin{figure}[htbp]
	 \centering	
				\includegraphics[scale=0.4]{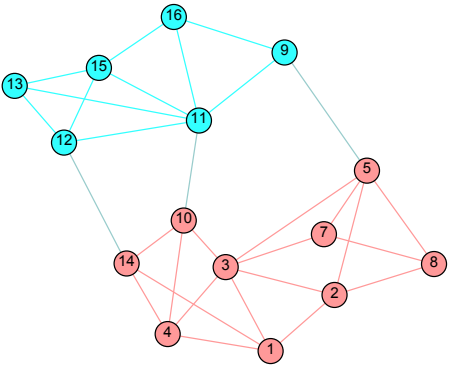}
				\caption{The example network G(2) with n = 15 in which node 6 was removed and node 16 was born. The green group dissolved and its members merged into blue and red groups. Colors represent communities discovered by \textit{LabelRankT}.}
				\label{fig:G2}			
\end{figure}		

The running time of the generalized LabelRank is the same as the original algorithm, as it is $O(m)$, linear with the number of edges $m$. The space complexity is $O(n)$ in practice because the number of labels in each node monotonically decreases and drops to a small constant in a few steps due to both cutoff and inflation operators. The $P$ matrix is replaced by sparse matrix representation of $n$ variable-length list of pairs (usually short) carried by each node; each pair contains label and its probability (with labels whose probability reduced to $0$ not listed).

Unlike in binary networks, there are various ways of adding a self-loop to each node to stabilize the results. It is interesting to see that the algorithm might perform slightly differently with different ways of defining self-loop. The most common ways include setting $w_{ii} =1$, $w_{ii} = max(w_{ik})$ or $w_{ii} = \sum_{k}{w_{ik}}$. In the experiments run for this paper, we use $w_{ii} =1$. Yet, it is still an open question how to optimally select the selfloop weight for each node.

\section{LabelRankT: An Extension for Dynamic Networks}
The extended algorithm called \textit{LabelRankT} is based on the generalized \textit{LabelRank} introduced in Alg.~\ref{alg1}. The description of \textit{LabelRankT} is contained in Alg.~\ref{alg2}. The main idea is to adjust our detection as the network structure changes. We take advantage of what we already obtain in previous snapshot for inferring the dynamics in the current time step. Since local structure information is encoded in the node label distributions, the evolving of communities is expected to be caught and reflected in these distributions.

\textit{LabelRankT} can be viewed as a \textit{LabelRank} with one extra conditional update rule by which only nodes involved any change accept the new distribution. Moreover, we only need to update nodes that are changed between two consecutive snapshots, including cases where an existing node adds or deletes links, or a node is removed from or newly joins the network. An example that shows different evolution events in three consecutive snapshots, G(0), G(1) and G(2) is shown in Figs. \ref{fig:G0}, \ref{fig:G1} and \ref{fig:G2}. During the evolution, nodes (edges) are added or removed, and communities split, merge and dissolve, all of which is captured by \textit{LabelRankT}. To discover communities in these snapshots, we ran our algorithm with the same parameters for all three snapshots.

In our algorithm, all these cases are handled by simply comparing neighbors of a node $i$ at two consecutive steps, $t-1$ and $t$, i.e., $Nb^t(i)$ and $Nb^{t-1}(i)$.  If $Nb^t(i)$ and $Nb^{t-1}(i)$ are not equal, then node $i$ is called a \textit{changed} node\footnote{Since it is often a case in practice, we assume here that all nodes in all steps are uniquely and consistently named.}. For changed nodes, we reinitialize their label distributions (i.e., $P^t$) and update until the simulation stops as in \textit{LabelRank}. However, since only changed nodes and their neighbors are involved (some neighbors only propagate labels but not update), \textit{LabelRankT} is more efficient than \textit{LabelRank}.

Such conditional update rule makes \textit{LabelRankT} applicable to dynamic networks for which changes arrive as a stream. When a new edge arrives in the incoming stream, \textit{LabelRankT} only updates the nodes that are attached to this edge. Thus, \textit{LabelRankT} can efficiently update and monitor communities whose evolution generates huge real-time data streams, such as the Internet and on-line social networks.

The time complexity for \textit{LabelRankT} can be derived as follows. It is easy to see for step (a) that we can track the number of changed nodes in O(1). For steps (b) and (c), since at most $n$ can be changed, we need to communicate across each edge at most twice in each iteration. Since the number of iterations required is a constant $T$ (usually less than 50 iterations), the overall complexity for detecting evolving communities between two consecutive snapshots is $O(Tm)$, implying $O(m)$ in general.
\begin{figure}
\centering	
				\includegraphics[scale=0.6]{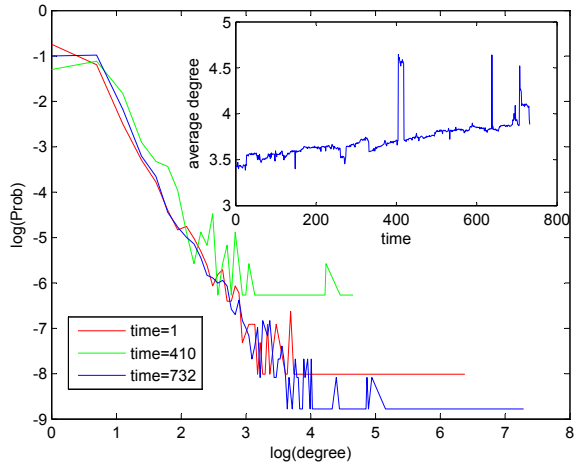}
				\caption{AS-Internet Routers Graph. Degree distributions at the beginning, middle and the end of evolution (main plot). Average degree over time (inset).}
				\label{fig:AS-1days-stat-deg}

\end{figure}
\begin{figure}
 \centering	
	 		  \includegraphics[scale=0.55]{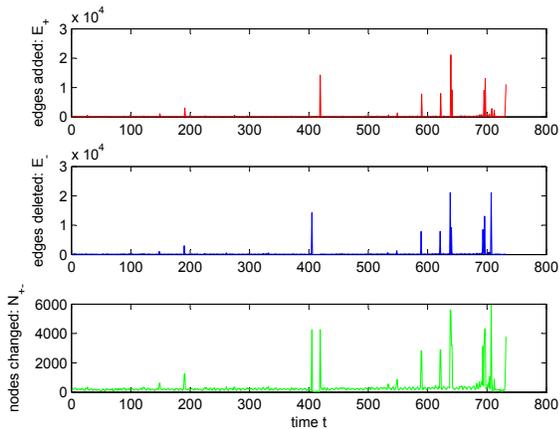}
				\caption{The structure changes in AS-Internet Routers Graph, including the number of edges added ($E_+$) and deleted ($E_-$), as well as the number of nodes involved in changes ($N_{+-}$).}
				\label{fig:AS-change}			
\end{figure}	

%-------------------------------------------
\section{Algorithm Performance Evaluation}
\label{sec:evaLabelRankT}
We tested the detection quality of \textit{LabelRankT} in terms of modularity and the efficiency in two real-world datasets.

%----------------------------------------------

\textbf{AS-Internet Routers Graph} \cite{Leskovecdyn:2005}. This is a communication network of who-talks-to-whom from the Border Gateway Protocol logs of routers in the Internet. The dataset contains 733 daily snapshots for 785 days from November 8 1997 to January 2 2000. The number of nodes in the largest snapshot is 6,477 (with 13,233 edges). The nodes and edges are added or removed over time. The structure at each snapshot could change dramatically as indicated by fluctuations in the average degree in Fig.~\ref{fig:AS-1days-stat-deg} and structure change statistics in Fig.~\ref{fig:AS-change}.

\textbf{arXiv HEP-TH}. High energy physics theory citation graph is from the e-print arXiv and covers all the citations within a dataset of 27,769 papers with 352,285 directed edges, each indicating that a paper at its tail cites the paper at its head.  The data covers papers from January 1993 to April 2003 \cite{Leskovecdyn:2005}. Unlike AS networks, it grows over time (i.e., edges and nodes are added but not removed).

\begin{figure*}
\centering	
	 \begin{minipage}[tb]{0.45\linewidth}
	 \centering	
				\includegraphics[scale=0.6]{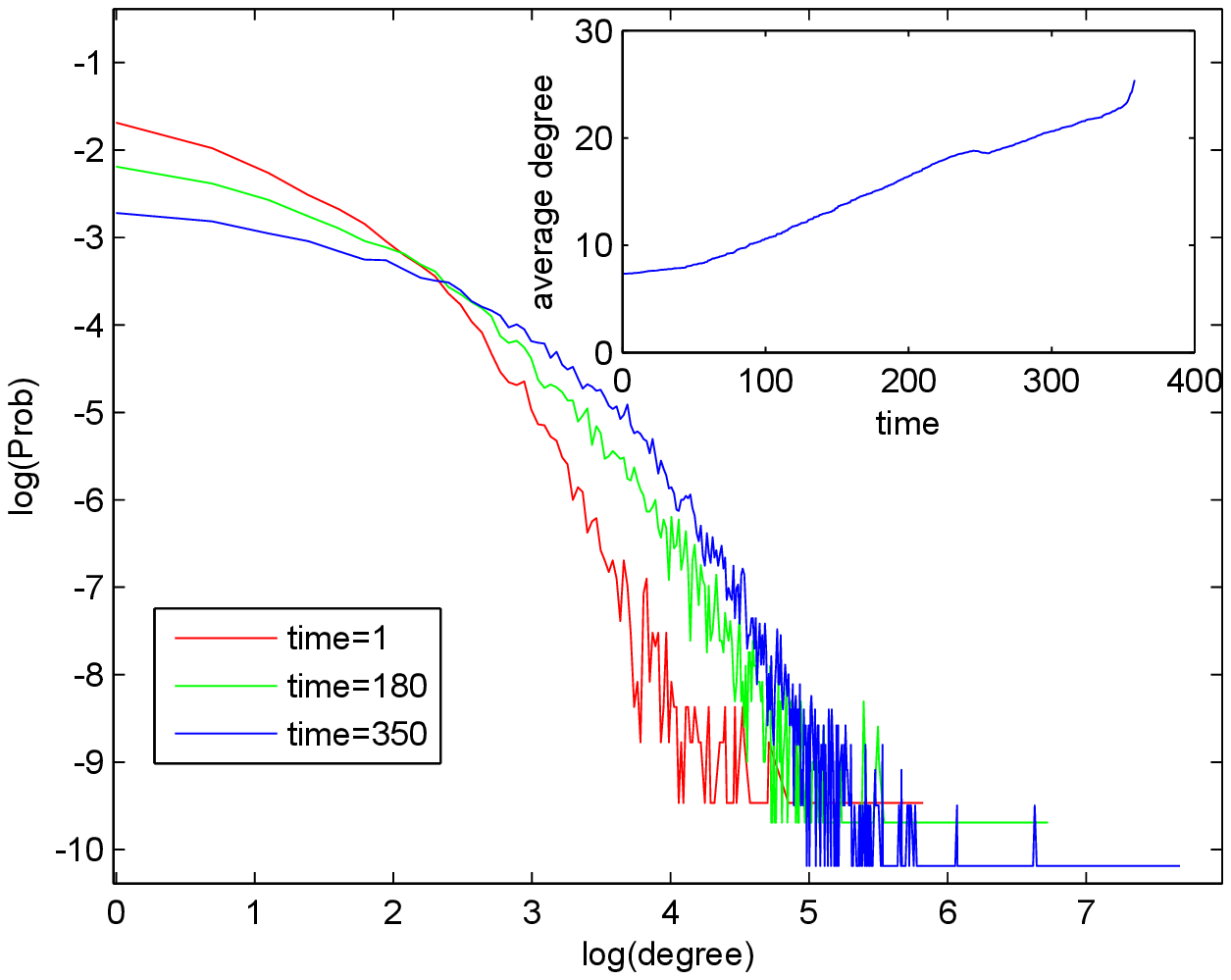}
				\caption{arXiv HEP-TH. Degree distributions at the beginning, middle and the end of evolution (main). Average degree over time (inset).}
				\label{fig:Hep-7days-stat-deg}
	 \end{minipage}	%	BECAREFULL, if you leave a empty line, then it is not side-by-side
	 \hspace{0.7cm}  %***
	 \begin{minipage}[tb]{0.45\linewidth}
	 \centering	
	 			\includegraphics[scale=0.55]{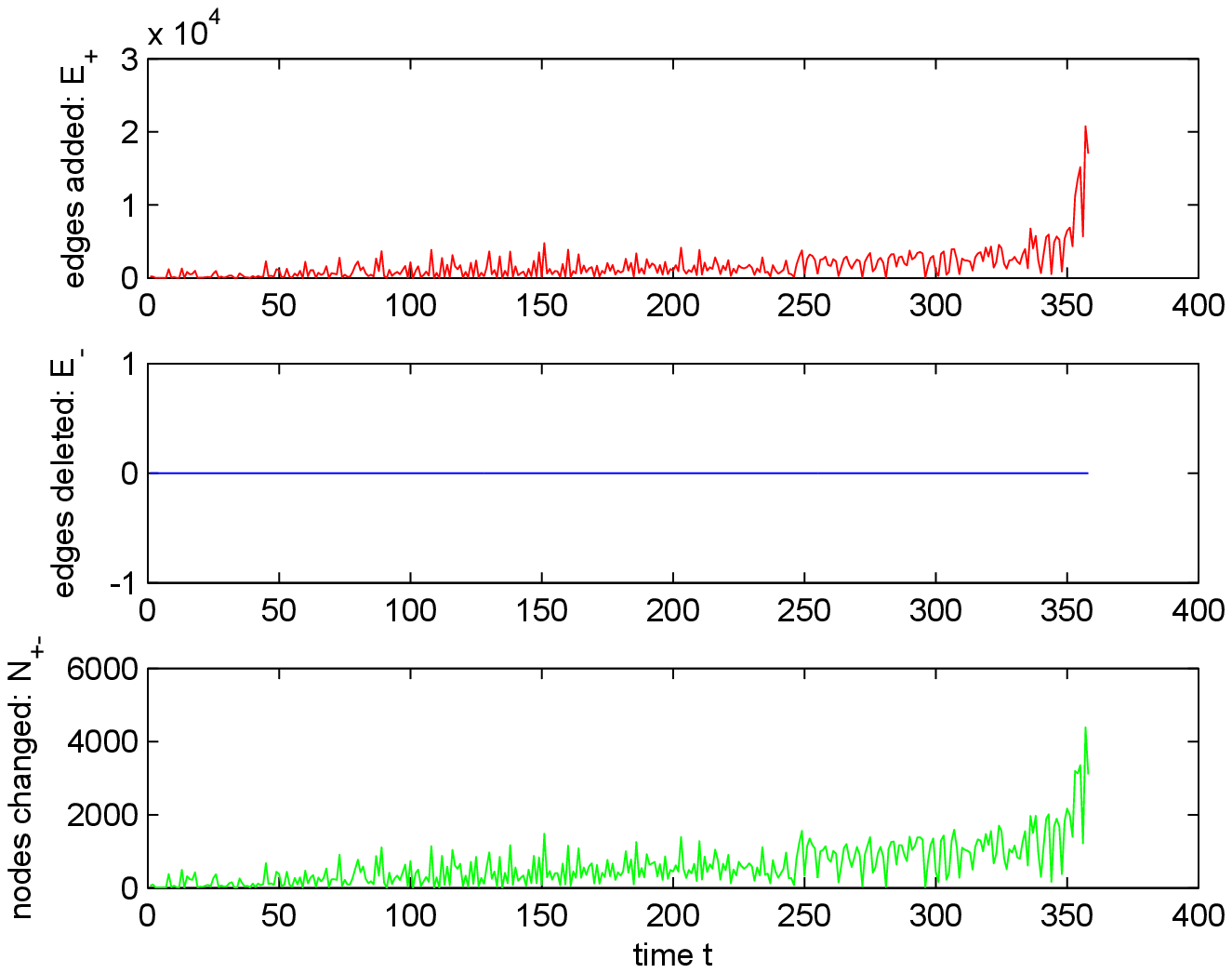}
				\caption{The structure changes in arXiv HEP-TH, including the number of edges added ($E_+$) and deleted ($E_-$), as well as the number of nodes involved in changes ($N_{+-}$).}
				\label{fig:Hep-7days-change}						
	\end{minipage}
\end{figure*}

%In our experiment, we made edges undirected.
The dataset is separated into snapshots by week (a total of 359 snapshots).
%The first snapshot consists of papers published before 1993.
The number of nodes in snapshots ranges from  12,917 to 27,769, while the number of edges ranges from 47,454 to 352,285. The structures over time are similar with a monotonic increase in average degree varying from 8 to 26 (see Fig.~\ref{fig:Hep-7days-stat-deg}). The statistics of structure change over time are shown in  Fig.~\ref{fig:Hep-7days-change}.
%\begin{figure*}[htbp]

\begin{figure*}[htbp]
\centering	
	 \begin{minipage}[t]{0.45\linewidth}
	 \centering	
				\includegraphics[scale=0.55]{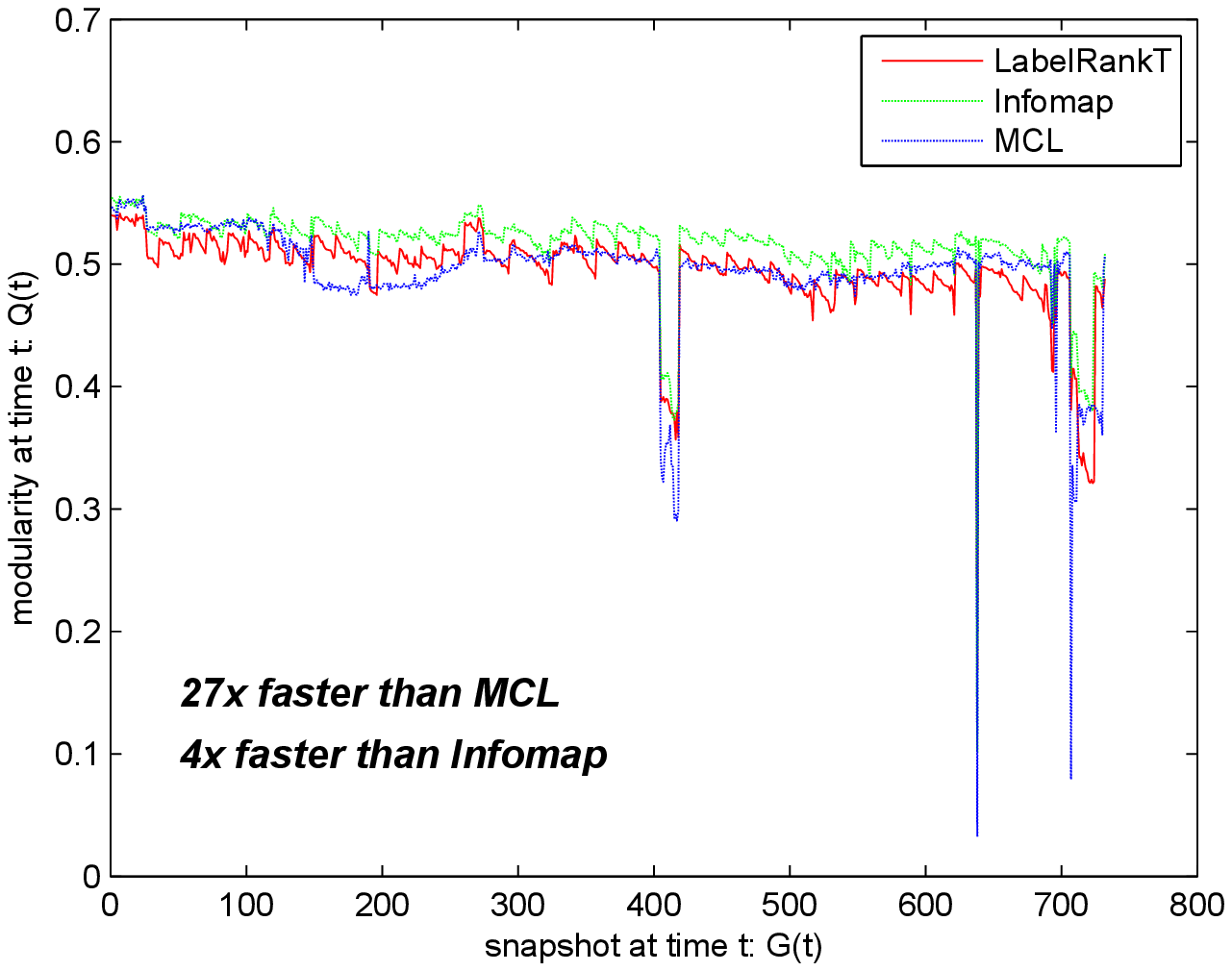}
				\caption{Comparison of modularity over time $Q(t)$ with static detection algorithms on AS-Internet Routers Graph.}
				\label{fig:AS-compare-static}
	 \end{minipage}	%	BECAREFULL, if you leave a empty line, then it is not side-by-side
	 \hspace{0.7cm}  %***
	 \begin{minipage}[t]{0.45\linewidth}
	 \centering	
				\includegraphics[scale=0.55]{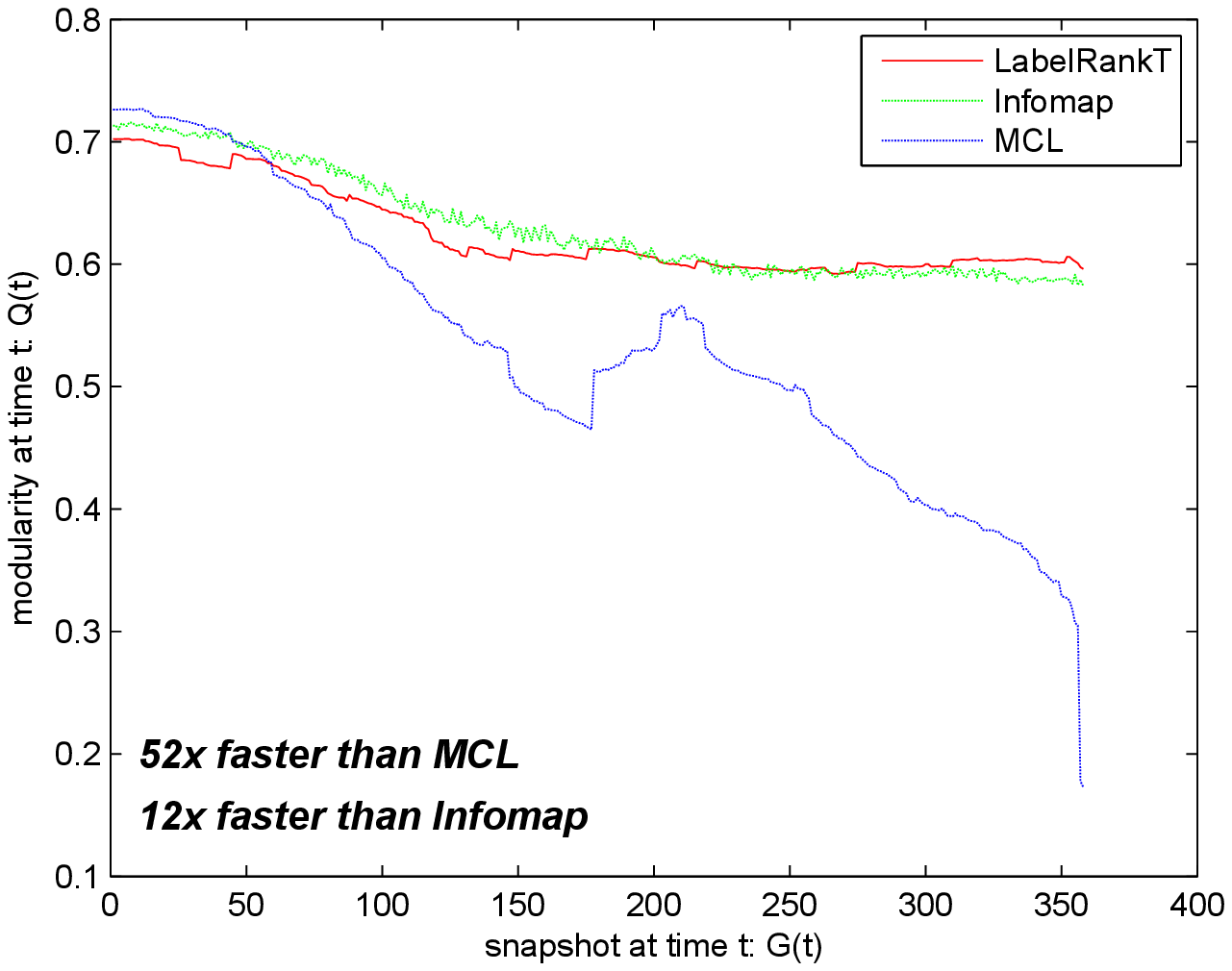}
				\caption{Comparison of modularity over time $Q(t)$ with static detection algorithms on arXiv HEP-TH.}
				\label{fig:Hep-7days-compare-static}
	 \end{minipage}	%	BECAREFULL, if you leave a empty line, then it is not side-by-side
\end{figure*}	
%----------------------------------------------
%side by side 1-2
%\begin{figure*}[htbp]
\begin{figure*}[htpb]
\centering	
	 \begin{minipage}[t]{0.45\linewidth}
	 \centering	
				\includegraphics[scale=0.55]{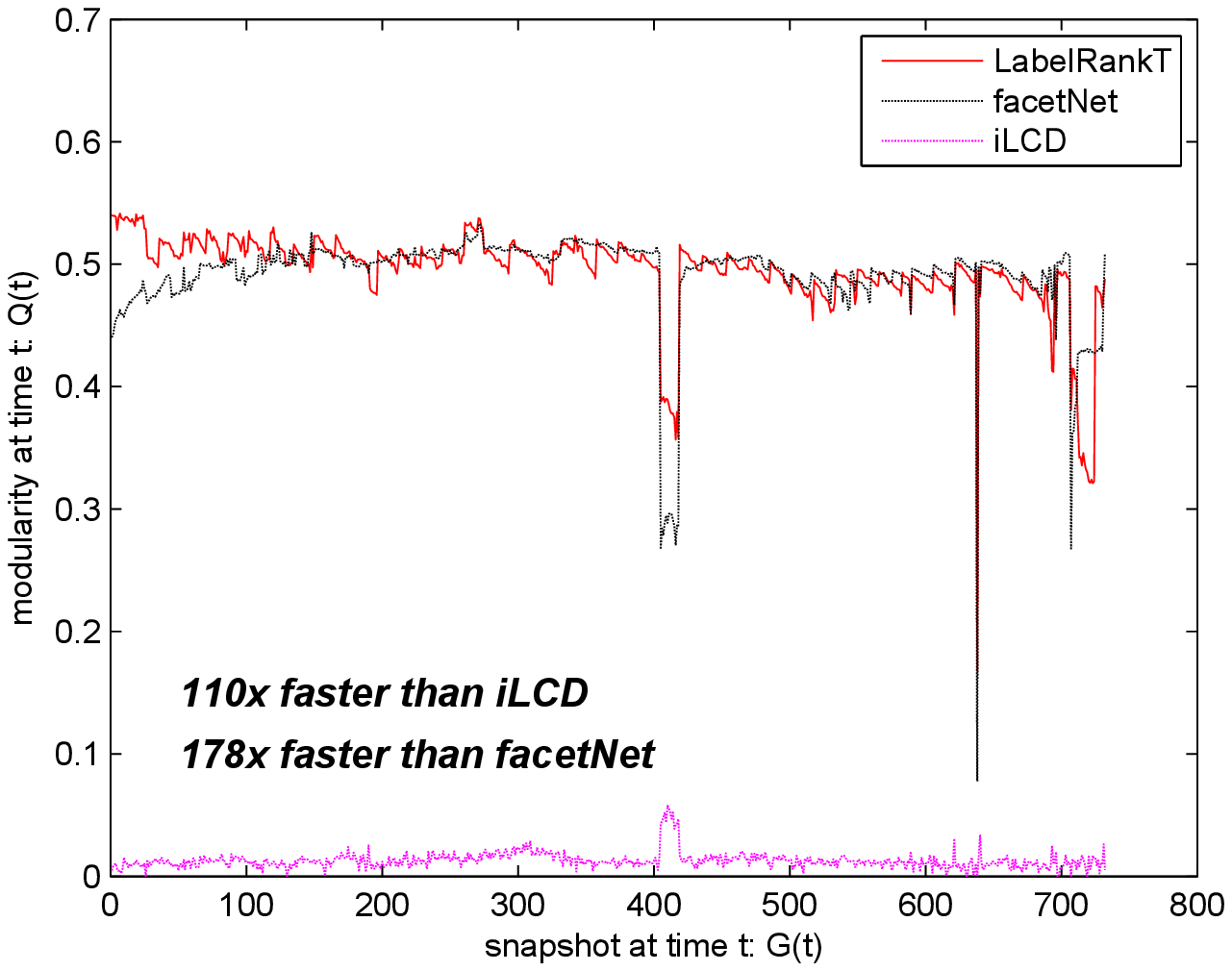}
				\caption{Comparison of modularity over time $Q(t)$ with dynamic detection algorithms on AS-Internet Routers Graph.}
				\label{fig:AS-compare-temporal}
	\end{minipage}
	 \hspace{0.7cm}  %***
	 \begin{minipage}[t]{0.45\linewidth}
	 \centering	
	 		\includegraphics[scale=0.55]{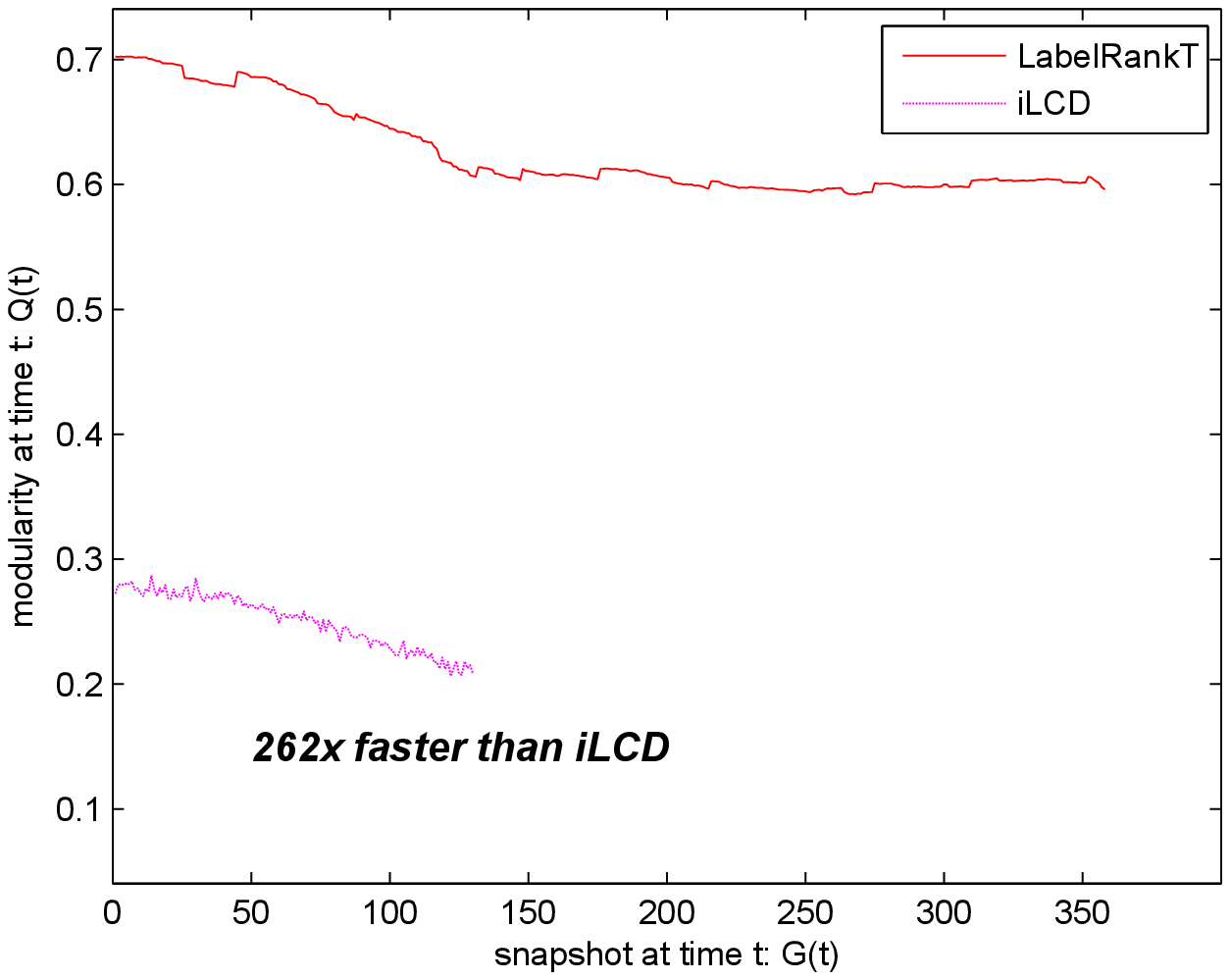}
			\caption{Comparison of modularity over time $Q(t)$ with dynamic detection algorithms on arXiv HEP-TH. We ran iLCD  on only the first 130 snapshots due to the time complexity. }
			\label{fig:Hep-7days-compare-temporal}
		\end{minipage}
\end{figure*}		
%----------------------------------------------
%side by side 1-2
%\begin{figure*}[htbp]
We first compared the performance of \textit{LabelRankT} with static algorithms MCL, using for all snapshots parameters that optimized performance for the first snapshot, and Infomap. Both run through each snapshot independently. Since a dynamic (especially incremental) algorithm like \textit{LabelRankT} does not recompute the entire network, static algorithms might perform better. In fact, on AS Graph, see Fig.~\ref{fig:AS-compare-static}, three algorithms actually have close performance. Infomap slightly outperforms \textit{LabelRankT} by about 5.03\% in modularity on average, \textit{LabelRankT} and MCL performance differs just by 0.43\%. On arXiv HEP-TH (which is of much larger size than AS), as seen in Fig.~\ref{fig:Hep-7days-compare-static}, Infomap and \textit{LabelRankT} perform within 0.88\% of each other. However, \textit{LabelRankT} outperforms MCL significantly by 15.37\% (Note that the behavior of MCL is partially caused by its sensitivity to parameters). On the other hand, \textit{LabelRankT} has benefit of efficiency. It runs 4 and 12 times faster than Infomap on AS Graph and arXiv HEP-TH respectively. And it is faster than MCL by a factor of 27 to 52 on AS Graph and arXiv HEP-TH, respectively.

%----------------------------------------------
%side by side 1-2
%\begin{figure*}[htbp]
\begin{figure*}[htpb]
\centering	
	 \begin{minipage}[tb]{0.45\linewidth}
	 \centering	
				\includegraphics[scale=0.55]{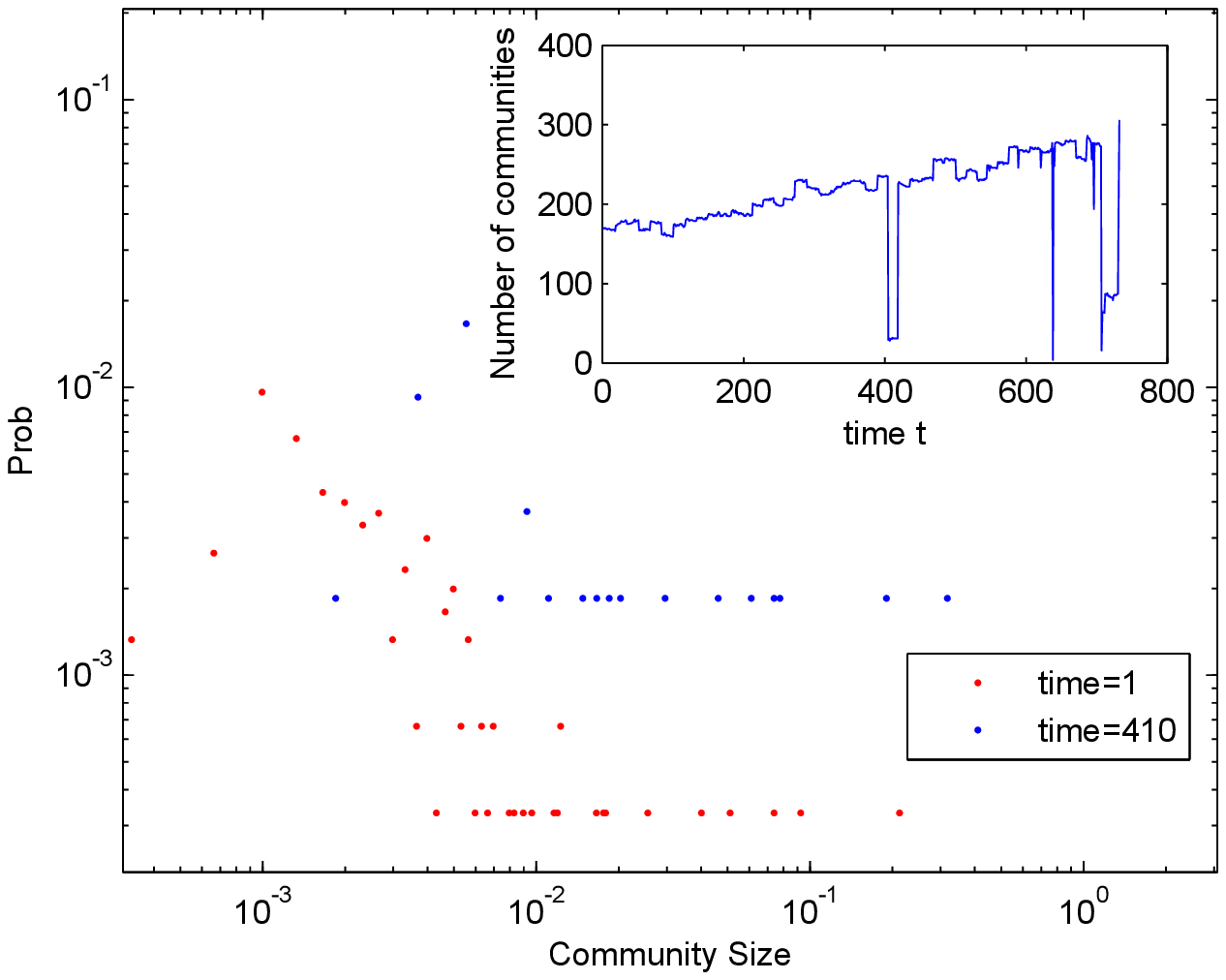}
				\caption{The community size distribution of AS-Internet Routers Graph tracked by \textit{LabelRankT} (loglog plot). Results at time 1 and 410 (dramatic changes occur) are shown in the main plot. The inset shows the number of communities over time.}
				\label{fig:AS-1days-comdist-bar-winsert}
	 \end{minipage}	%	BECAREFULL, if you leave a empty line, then it is not side-by-side
	 \hspace{0.7cm}  %***
	 \begin{minipage}[tb]{0.45\linewidth}
	 \centering	
				\includegraphics[scale=0.45]{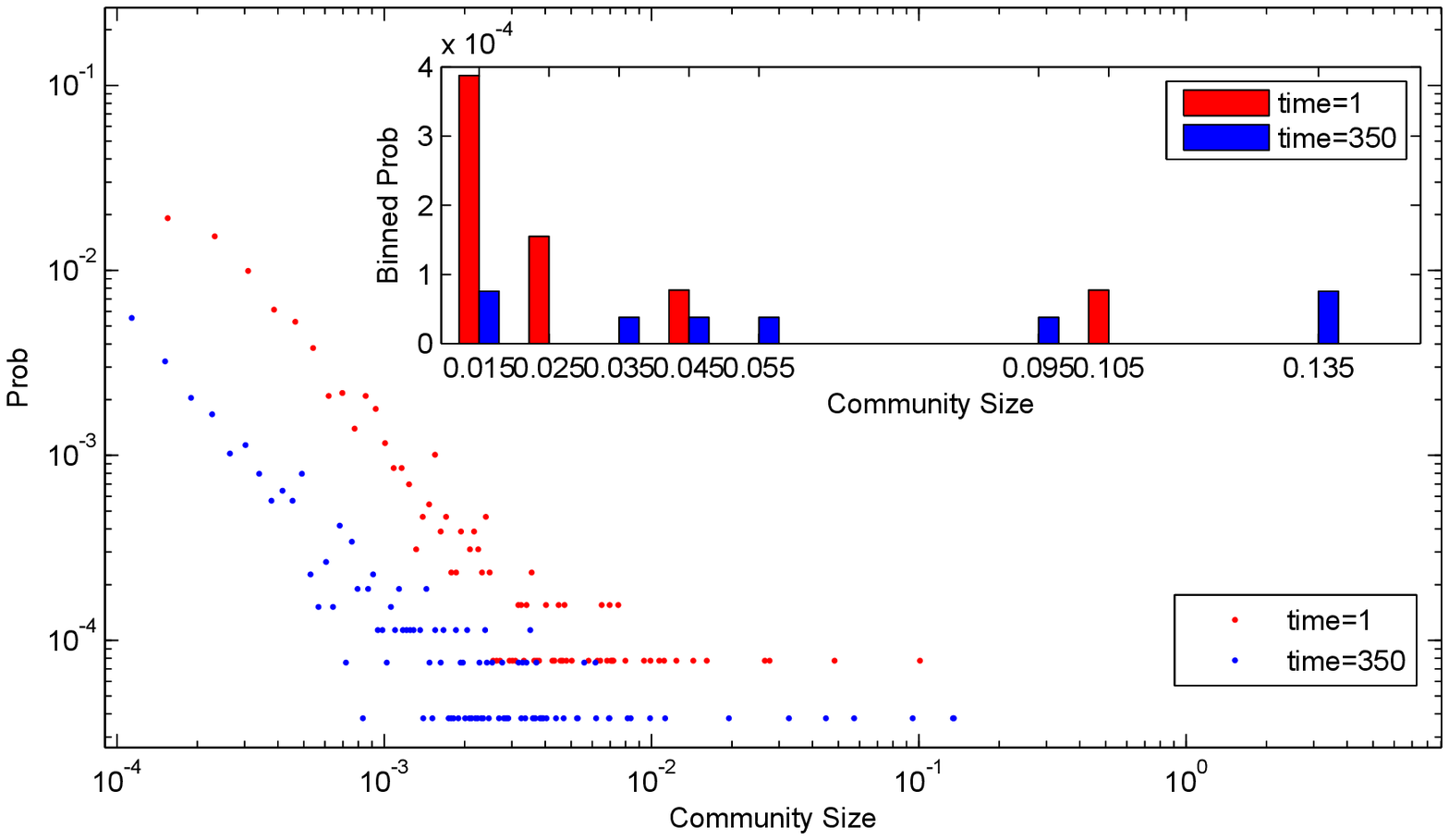}
				\caption{The community size distribution of arXiv HEP-TH  tracked by \textit{LabelRankT} (loglog plot). Results at time 1 and 350 (near the end of evolution) are shown. In the inset, the tails of the distributions in main plot are binned with a width 0.01 to show the shift in large size communities.}
				\label{fig:Hep-7days-comdist-bar-winsert}	
	\end{minipage}
\end{figure*}	
We also compared \textit{LabelRankT} with two publicly available dynamic algorithms that employ incremental detection methods : facetNet \footnote{Since facetNet requires the number of communities as input, we assign it the value produced by \textit{LabelRankT}.} \cite{facetNet:2009} and iLCD \footnote{After detection, if a node belongs to more than one community, we assign it to the the one with maximum size to be able to output only unique and disjoint partitions.} \cite{iLCDCazabet:2010}. On AS Graph, see Fig.~\ref{fig:AS-compare-temporal}, facetNet and \textit{LabelRankT} achieve performance within just 0.07\% of each other, while iLCD fails to find strong community structure at all. As shown in Fig.~\ref{fig:Hep-7days-compare-temporal}, on arXiv HEP-TH, facetNet does not work due to the overflow in memory, while \textit{LabelRankT} performs at least twice better than iLCD. Moreover, \textit{LabelRankT} is more than 100 times faster than both facetNet and iLCD on the two datasets used here.

We also analyzed the number of communities and the distribution of community sizes relative to $n$ (i.e., the probability of seeing a community with certain size), produced by \textit{LabelRankT}. As shown in Fig.~\ref{fig:AS-1days-comdist-bar-winsert}, AS Graph does not evolve smoothly all the time. The abrupt drop in the number of communities at time 410 signals a dramatic change in structure, which is verified by a completely different distribution of community sizes in comparison with the beginning one. Although this violates our assumption, \textit{LabelRankT} still worked well as evidenced by consistency of its results with the results of static algorithm Infomap. In contrast, arXiv HEP-TH exhibits a fairly smooth pattern shown in Fig.~\ref{fig:Hep-7days-comdist-bar-winsert}. The distributions of community sizes at time 1 and 350 (near the end of evolution) obey power laws with essentially identical exponents. Small size communities grow faster as more and more papers are published as indicated by the downward shift in these distributions. Some communities grow relatively faster than the others and the largest communities expand as indicated by the shift to the right (see the inset).

\section{Effect of Edge Weight and Direction}
In this section, we demonstrate that incorporating the edge weight and direction into the network description allows \textit{LabelRankT} to identify communities better. The experiments were conducted on a dataset including weighted and directed networks.

%-------------------------------table-------------------------------------
\begin{table*}
\caption{The \textit{average modularity differences} between the weighted and directed version and the unweighted and undirected version of \textit{LabelRankT} with different values of the conditional update parameter $q$ on all the $43$ snapshots of the Reality Mining Bluetooth Scan data.}
\label{t:bluetooth}
\centering
\begin{tabular}{c||c|c|c|c|c|c|c|c|c|c|c}
\hline \hline
     Conditional update parameter $q$ & 0.05 & 0.1 & 0.2 &	0.3 & 0.4 &	0.5 & 0.6 &	0.7 & 0.8 &	0.9 & 0.95 \\
\hline
     \textit{Average modularity difference} & 0.101 & 0.13 & 0.169 & 0.174 &	0.181 & 0.197 & \textbf{0.206} & 0.201 & 0.199 & 0.193 & 0.19 \\
\hline \hline
\end{tabular}
%\vspace{-1.5em}
\end{table*}
%---------------------------------table------------------------------------

\textbf{Reality Mining Bluetooth Scan Data} \cite{RealityMining}. This dataset was created from the records of Bluetooth Scans generated among the $94$ subjects in Reality Mining study conducted from 2004-2005 at the MIT Media Laboratory. In the network, nodes represent the subjects and the directed edges correspond to the Bluetooth Scan records and the weight of each edge represent the number of directed Bluetooth scans between the two subjects. In the comparison described below, we only adopted the records from August 02, 2004 (Monday) to May 29, 2005 (Sunday) and we divided them weekly snapshots, so each snapshot represents scans collected during the corresponding week. There are total of 43 snapshots.

\begin{figure}	
  \centering
	\includegraphics[scale=0.6]{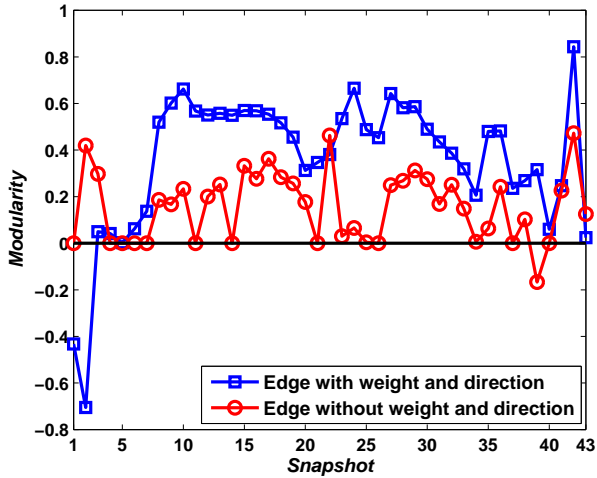}
	\caption{The modularity of \textit{LabelRankT} with and without edge weight and direction on Reality Mining Bluetooth Scan data with the conditional update parameter $q=0.6$.}
	\label{fig:modularity_diff}
\end{figure}

We compared the community detection results produced by \textit{LabelRankT} on the Reality Mining Bluetooth Scan network with and without edge weight and direction. By varying the parameter $q$ (from $0.05$ to $0.95$) of the conditional update $\Theta _{q}$ operator, we calculated the \textit{average modularity differences}, shown in Table~\ref{t:bluetooth}, between the weighted and directed version and the unweighted and undirected version of \textit{LabelRankT}. All the \textit{average modularity differences} in Table~\ref{t:bluetooth} are positive. This demonstrates that including the edge weight and direction improves the performance of our algorithm. Further, Fig.~\ref{fig:modularity_diff} presents the modularity of \textit{LabelRankT} with and without edge weight and direction on all the $43$ snapshots with the conditional update parameter $q=0.6$ when the \textit{average modularity difference} is the largest. In conclusion,
these results demonstrate that inclusion of edge weight and direction of the network improves the quality of communities detected by \textit{LabelRankT}.

\section{Related Work}
Label propagation and random walk based algorithms are most relevant to our work.
LPA \cite{Raghavan:2007,JieruiXieLPA:2010} identifies disjoint groups as nodes with the same label. COPRA \cite{Gregory:2010} and SLPA \cite{JieruiXie-SLPA:2012} extend LPA to detection of overlapping communities by allowing multiple labels. However, none of these algorithm resolves the LPA randomness issue, where different communities may be detected in different runs over the same network. Markov Cluster Algorithm (MCL) proposed in \cite{MCL:2000} is based on simulations of flow (random walk). MCL executes repeatedly matrix multiplication followed by inflation operator.

\textit{LabelRankT}, like its predecessor \textit{LabelRank} (see \cite{JieruiXie-LabelRank-NSW:2013}) differs from MCL in at least two aspects. First, \textit{LabelRankT} applies the inflation to the label distributions and not to the matrix $M$. Second, the update of label distributions on each node in \textit{LabelRankT} requires only local information. Thus it can be computed in a decentralized way. Regularized-MCL \cite{kdd-2009-R-MCL} also employs a local update rule of label propagation operator. Despite that, the authors observed that it still suffers from the scalability issue of the original MCL. To remedy, they introduced Multi-level Regularized MCL, making it complex. In contrast, we address the scalability by introducing new operator, conditional update, and the novel stopping criterion, preserving the speed and simplicity of the LPA based algorithms. Moreover, neither MCL nor Regularized-MCL is suitable for dynamic networks.

For dynamic networks, there has been work that focus on exploring the properties of evolving communities that could be used to guide the detection algorithms. Palla et al. \cite{Palla:2007} developed an algorithm based on the clique percolation method and investigated the time dependence of overlapping communities to uncover basic relationships characterizing community evolution. Tantipathananandh and Berger-Wolf \cite{SocialCostModel} extended their previous social cost model to arbitrary dynamic networks and approximately solved the optimization problem using semidefinite programming relaxation and a rounding heuristic. Bassett et al. \cite{RobustCommunityDetection} proposed an approach to construct representative partitions. This approach adopts a null model to correct for statistical noise in sets of partitions to improve robustness of community detection result in time-dependent networks. 

Typically, an incremental detection \cite{Aynaud:2011} considers the stream of changes between snapshots explicitly as opposed to applying static algorithms to each static snapshot \cite{BergerWolf:2006, Asur:2009}. Ning et al. \cite{Ning:2007} proposed an incremental spectral clustering that continuously updates the eigenvalues and eigenvectors by computing approximations of the generalized eigenvalue system of the normalized cut. facetNet \cite{facetNet:2009} extends the non-negative matrix factorization algorithm. Its drawback is that it requires the number of communities to be provided as an input. GraphScope \cite{GraphScope:2007} is a parameter-free algorithm where the minimum description length principle is
used to extract communities and to detect the changes. It does not consider the deletion of nodes. Bansal et al. \cite{Bansal:2010} extended CNM algorithm. However, their approach is limited to networks that change very little from snapshot to snapshot. Similarly, Gorke et al. \cite{Robert:2010} modified both global and local modularity optimization based CNM algorithm and Louvain algorithm, but this modification can handle only small changes between snapshots. iLCD \cite{iLCDCazabet:2010} updates the existing
community by adding a new node to it if the node's number of second neighbors and number of robust second neighbors are greater than expected values. A limitation is that it can not add two new nodes and a link in-between them at the same time. 

\section{Conclusions}
Aiming at a highly efficient and general on-line detection algorithm, we introduce \textit{LabelRankT} for incremental detection of evolving communities in large-scale dynamic networks through label propagation. \textit{LabelRankT} is based on the introduced generalized \textit{LabelRank}, in which each node requires only local information during label propagation processing. \textit{LabelRankT} is also able to detect communities in various network types including networks with directed/undirected and weighted/unweighted edges in linear time.

In future work, we plan to apply \textit{LabelRankT} to self-organizing applications such as ad-hoc mobile networks and P2P networks, where each node corresponds to a physical platform. By taking into account temporal and spatial (communities) correlations, we will attempt to construct efficient distributed social-based message routing algorithm on top of \textit{LabelRankT}. We also plan to extend \textit{LabelRankT} to overlapping community detection \cite{JieruiXie-Survey:2013}.

%\vspace{1cm}
\section*{Acknowledgment}

This work was supported in part by the Army Research
Laboratory under Cooperative Agreement Number
W911NF-09-2-0053 and by the Office of Naval Research
Grant No. N00014-09-1-0607. The views and
conclusions contained in this document are those of the
authors and should not be interpreted as representing
the official policies either expressed or implied of the
Army Research Laboratory, the Office of Naval Research, or the U.S. Government.

\bibliographystyle{abbrv}

%\bibliography{CommunityBIB-Jerry}  % sigproc.bib is the name of the Bibliography in this case!

\end{document}